\documentclass[twocolumn,british,sort&compress]{article}
\usepackage[T1]{fontenc}
\usepackage[latin9]{inputenc}
\usepackage[a4paper]{geometry}
\usepackage{color}
\usepackage{babel}

\usepackage{amsmath}
\usepackage{graphicx}
\usepackage{esint}
\usepackage[numbers]{natbib}
\usepackage[unicode=true, pdfusetitle,
 bookmarks=true,bookmarksnumbered=false,bookmarksopen=false,
 breaklinks=false,pdfborder={0 0 1},backref=false,colorlinks=true]
 {hyperref}
\hypersetup{
 citecolor=blue,urlcolor=blue}
\begin{document}
\global\long\def\mat#1#2#3{#1_{\hphantom{#2}#3}^{#2}}

\twocolumn[\begin{@twocolumnfalse}

\title{Non-local approach to kinetic effects on parallel transport in fluid
models of the scrape-off layer}

\author{John Omotani and Ben Dudson\\
{\small York Plasma Institute, Department of Physics, University
of York, Heslington, York, YO10 5DD, UK}\\
{\small Email: john.omotani@york.ac.uk}}

\date{}
\maketitle
\begin{abstract}
By using a non-local model, fluid simulations can capture kinetic
effects in the parallel electron heat-flux better than is possible
using flux limiters in the usual diffusive models. Non-local and diffusive
models are compared using a test case representative of an ELM crash
in the JET SOL, simulated in one dimension. The non-local model shows
substantially enhanced electron temperature gradients, which cannot
be achieved using a flux limiter. The performance of the implementation,
in the BOUT++ framework, is also analysed to demonstrate its suitability
for application in three-dimensional simulations of turbulent transport
in the SOL.

\end{abstract}
\end{@twocolumnfalse}]

The divertor target heat-flux will be an important limit on the performance
of future magnetic confinement fusion devices, from ITER onwards\citep{Loarte2007_ITER,loarte2010iter}.
Modelling of transport in the scrape-off layer (SOL) is therefore
a subject of urgent interest. The conditions in the SOL---large amplitude
fluctuations, open magnetic field lines---prevent the application
of approximations that are used to make (gyro-)kinetic simulations
of turbulence feasible in the core plasma; kinetic simulations are
limited to one-dimensional models (e.g.~\citep{pittsJET_ELM}). Nevertheless,
three-dimensional features---turbulence and {}`blob'\citep{d'ippolito:060501}
motion---are critical parts of the picture, especially in determining
the width of the scrape-off layer\citep{myra:012305,Militello_MAST_ESEL}.
These are studied using fluid simulations which are either two-dimensional
(e.g.~ESEL\citep{PhysRevLett.92.165003,garcia:082309}, RI/FI-SOL\citep{yu:042508},
SOLT\citep{russell:122304} and G-ESEL\citep{madsen:112504}) or use
simple (Braginskii) models for the parallel dynamics (e.g. heat diffusion
in GBS\citep{Ricci_GBS} or Spitzer conductivity in BOUT++\citep{PhysRevLett.108.215002,Angus_3dblob_BOUT++}).
However, such simple treatments of the parallel transport may not
give the full picture, as at typical ELM parameters the electron collision
length exceeds the total connection length between the divertor plates:
for example, at a density of $1.0\times10^{19}\textrm{m}^{-3}$ and
an electron temperature of 300eV, the electron collision length is
$\sim110\textrm{m}$ while the connection length in JET is only $\sim80\textrm{m}$.
Furthermore, as a consequence of an edge localized mode (ELM) or the
motion of a blob through the SOL there will be large changes in the
overall density and temperature on a field line. When such transient
events occur kinetic simulations show that the diffusive description
is not a good model of the parallel transport: the correction for
{}`kinetic effects' that can be applied by using flux limiters breaks
down since the appropriate limiter, as determined by comparison with
kinetic simulations, varies strongly in time\citep{Tskhakaya2008}.

To enable a self-consistent treatment of turbulence with {}`kinetically-corrected'
parallel transport, new methods for including kinetic effects in fluid
models are required. This paper describes the application of a non-local
calculation of the electron heat-flux\citep{ji:022312}, which has
not previously been applied to the SOL. The parallel electron heat-flux
is derived by using a very high order truncation (with typically 100
to 1600 moments retained) to solve a one-dimensional reduced drift
kinetic equation in terms of a set of integrals. This model has a
couple of particularly attractive features. Firstly, being derived
directly from the kinetic equation, it has no ad-hoc parameters. Secondly,
a good deal of the computational complexity involved is found in solving
for the eigensystem that is used to decouple the moment equations.
This need be done only once for a given truncation, and so a substantial
piece of the {}`kinetic' part of the calculation is removed from
the simulations. Alternatives, which have been used in laser-plasma
physics\citep{luciani1983nonlocal,batishchev:2302}, employ integral
kernels which would be numerically very challenging when, as is the
case here, the density cannot be assumed to be constant.

The model has been implemented in the BOUT++ plasma fluid simulation
framework\citep{Dudson2009} in order to expedite its future inclusion
in three-dimensional simulations. Here we evaluate its performance
through one-dimensional simulations of parallel transport, with parameters
chosen to approximate an ELM in JET\citep{havlivckova2012comparison}. 

In Section \ref{sec:Theory} we outline the model and detail some
additions needed for its use in the SOL; in Section \ref{sec:Simulations}
the results of simulations of the test case are presented, comparing
the non-local model to the Braginskii model with and without flux
limiters; in Section \ref{sec:Numerical-Performance} the numerical
performance of the implementation is discussed, which is particularly
important for future extensions to three-dimensional fluid models;
finally in Section \ref{sec:Conclusions} we conclude and discuss
future developments.

\section{Theory\label{sec:Theory}}

\subsection{Non-local heat-flux\label{sub:Model}}

The thermal speed of the electrons in SOL plasmas is very much higher
than that of the ions. Hence in calculations of the electron dynamics,
we may consider the evolution of the background profiles to be slow.
Solving a static electron kinetic equation for the heat-flux will
therefore give a good approximation to its true value, which can then
be used in the evolution equations for the background fields.

The approach taken in \citep{ji:022312} starts from the one-dimensional
kinetic equation for the non-Maxwellian part $\delta f_{e}$ of the
electron distribution function whose Maxwellian part is $f_{e}^{(0)}$:\begin{align}
v_{\|}\frac{\partial\langle\delta f_{e}\rangle}{\partial\ell} & =\sum_{a=e,i}C\!\left(\langle f_{e}^{(0)}\!+\!\delta f_{e}\rangle,\langle f_{a}^{(0)}\!+\!\delta f_{a}\rangle\right)-v_{\|}\frac{\partial\langle f_{e}^{(0)}\rangle}{\partial\ell}\label{eq:kinetic_equation}\end{align}
where $\langle\cdot\rangle$ denotes the gyroaverage, $v_{\|}$ is
the component of the velocity parallel to the magnetic field, $\ell$
is the distance along the field line and $C\left(\cdot,\cdot\right)$
is the linearized Fokker-Planck collision operator. We expand in fluid
moments on a basis $P^{lk}(\frac{\mathbf{v}}{v_{Te}})=P^{l}(\frac{\mathbf{v}}{v_{Te}})L_{k}^{(l+\frac{1}{2})}(\frac{v^{2}}{v_{Te}^{2}})$,
$P^{l}(\frac{\mathbf{v}}{v_{Te}})$ being tensor harmonic polynomials
and $L_{k}^{(l+\frac{1}{2})}(\frac{v^{2}}{v_{Te}^{2}})$ associated
Laguerre polynomials (see \citep{ji2006exact} for details), and truncate
to $L$ angular harmonics and $K$ Laguerre orders, i.e.~$0\leq l<L$,
$0\leq k<K$; here we will take $L=20$, $K=20$ and so have 400 moments
in total. After truncating it is convenient to represent the pairs
$(l,k)$ with a single index, $A,B,\ldots$, running over $L\times K$
possible values. Then \eqref{eq:kinetic_equation} reduces to a set
of one-dimensional, first order ODEs for the non-Maxwellian fluid
moments\begin{align}
\sum_{B}\mat{\Psi}AB\frac{\partial n^{B}}{\partial z} & =\sum_{B}\mat CABn^{B}+g^{A}\label{eq:moment-equations}\end{align}
where $z$ is the dimensionless length defined by $\frac{\partial z}{\partial\ell}=\frac{1}{\Lambda_{C}}$
(with $\Lambda_{C}$ the electron-electron collision length); $n^{A}$
are the parallel fluid moments of $\delta f_{e}$, \begin{align}
\left\langle \delta f_{e}\right\rangle  & =\sum_{lk}\frac{e^{-v^{2}/v_{Te}^{2}}}{\pi^{\frac{3}{2}}v_{Te}^{3}}P_{l}\!\left(\frac{v_{\|}}{v_{Te}}\right)L_{k}^{(l+\frac{1}{2})}\!\left(\frac{v^{2}}{v_{Te}^{2}}\right)n^{(l,k)}\end{align}
where $P_{l}(\frac{v_{\|}}{v_{Te}})$ are Legendre Polynomials, coming
from $\left\langle P^{l}(\frac{\mathbf{v}}{v_{Te}})\right\rangle =P_{l}(\frac{v_{\|}}{v_{Te}})P^{l}(\frac{\mathbf{B}}{B})$;
we neglect $\frac{1}{T}\frac{\partial T}{\partial\ell}n^{A}$ on the
assumption that $\frac{n_{e}}{T}\frac{\partial T}{\partial\ell}$
is of the same order as $n^{A}$ so that their product is small; $\mat{\Psi}AB$
is the matrix coming from $v_{\|}$,\begin{align}
\mat{\Psi}{(l,k)}{(l',k')} & =\psi_{kk'}^{l}\delta_{l+1,l'}+\psi_{k'k}^{l-1}\delta_{l-1,l'}\\
\psi_{kk'}^{l} & =\frac{(l+1)}{(2l+1)(2l+3)}\lambda_{k'}^{l+1}(\delta_{k,k'}-\delta_{k-1,k'})\end{align}
where $\lambda_{k}^{l}=(l+k+\frac{1}{2})!/k!/(\frac{1}{2})!$; $\mat CAB$
is the (dimensionless) collision matrix\begin{align}
\mat C{(l,k)}{(l',k')} & =\frac{\Lambda_{C}}{n_{e}v_{Te}}\sigma_{l}\left(A_{ee}^{lkk'}+B_{ee}^{lkk'}+A_{ei}^{lkk'}\right)\delta_{l,l'}\end{align}
where $\sigma_{l}=\frac{l!(\frac{1}{2})!}{2^{l}(l+\frac{1}{2})!}$,
$A_{ee}^{lkk'}$ and $B_{ee}^{lkk'}$ are the moments of the electron-electron
collision operator and $A_{ei}^{lkk'}$ are the moments of the electron-ion
collision operator calculated in \citep{ji2006exact} ($B_{ei}^{lkk'}$
vanishes at lowest order in the electron-ion mass ratio); and \begin{align}
g^{A} & =\frac{5}{4}\frac{n_{e}}{T_{e}}\frac{\partial T_{e}}{\partial z}\delta_{A,(1,1)}\end{align}
 is the drive term from the gradient of the Maxwellian part, $f_{e}^{(0)}$.
These equations are valid only for the non-Maxwellian moments (i.e.
$A,B=(0,0),(0,1)\text{ and }(1,0)$ are excluded): the Maxwellian
moments (the density, temperature and fluid velocity) are background
fields which give the drive term $g^{A}$.

The equations \eqref{eq:moment-equations} can be decoupled using
the eigenvectors of the matrix operator $\mat{\left(C^{-1}\Psi\right)}AB$
to give\begin{align}
\zeta_{(A)}\frac{\partial\hat{n}^{A}}{\partial z} & =\hat{n}^{A}+\hat{g}^{A}\end{align}
where $\hat{n}^{A}$ and $\hat{g}^{A}$ are the components of $n^{A}$
and $g^{A}$ on the eigenvector basis and $\zeta_{(A)}$ are the corresponding
eigenvalues (note that as defined here $\zeta_{(A)}$ are the reciprocals
of the eigenvalues used in \citep{ji:022312}%
\footnote{The reason for this discrepancy is that in \citep{ji:022312} the
eigenvalues of $\mat{\left(\Psi^{-1}C\right)}AB$ are found. Due to
the block-off-diagonal structure of $\Psi$, when the three Maxwellian
moments are removed the upper-left-most blocks are no longer square,
with the result that $\Psi$ is not invertible. The authors of \citep{ji:022312}
resolved this by changing the truncation to keep three extra moments
$(0,K)$, $(0,K+1)$ and $(1,K)$ so that all the blocks of $\Psi$
remain square. Here we have not added these extra moments and so must
find the eigenvalues and eigenvectors of $\mat{\left(C^{-1}\Psi\right)}AB$
(since $C$ remains block diagonal and so is still invertible), and
must account for one of the eigenvalues being zero.%
}). The solutions of these equations can be expressed as\begin{align} \hat{n}^{A}(z) & =\begin{cases} -\hat{g}^{A} & \zeta_{(A)}=0\\[6pt] \begin{array}{l} \hat{n}^{A}(z_{0})\exp\left(\frac{z-z_{0}}{\zeta_{(A)}}\right)\\[4pt] +{\displaystyle\int_{z_{0}}^{z}}dz'\exp\left(\frac{z-z'}{\zeta_{(A)}}\right)\frac{\hat{g}^{A}(z')}{\zeta_{(A)}}\end{array} & \zeta_{(A)}\neq0\end{cases}\label{eq:moment-solution}\end{align}where
$\hat{n}^{A}(z_{0})$ are the values at the boundary $z_{0}$ (where
$z_{0}\equiv z_{-}<z$ for $\zeta_{(A)}<0$ and $z_{0}\equiv z_{+}>z$
for $\zeta_{(A)}>0$).

Finally, after transforming back to the original vector basis we use
the (1,1) moment to calculate the parallel heat-flux, which is needed
to close the fluid equations:\begin{align}
q_{e\|} & =-\frac{5}{4}v_{Te}T_{e}n^{(1,1)}=-\frac{5}{4}v_{Te}T_{e}\sum_{B}\mat W{(1,1)}B\hat{n}^{B}\end{align}
where $\mat WAB=W_{(B)}^{A}$ is the matrix formed from the eigenvectors
$W_{(B)}$ (whose eigenvalues are $\zeta_{(B)}$).

So to calculate $q_{e\|}$ we must (a) compute integrals over $z$
(as many as the number of moments retained) and (b) set the boundary
values $\hat{n}^{A}(z_{0})$ (which we will do so as to impose the
boundary condition on the heat-flux, Section \ref{sub:Boundary-Conditions}).

\subsection{Fluid Equations\label{sub:Fluid-Equations}}

\begin{figure*}[t]
\includegraphics[width=0.49\textwidth]{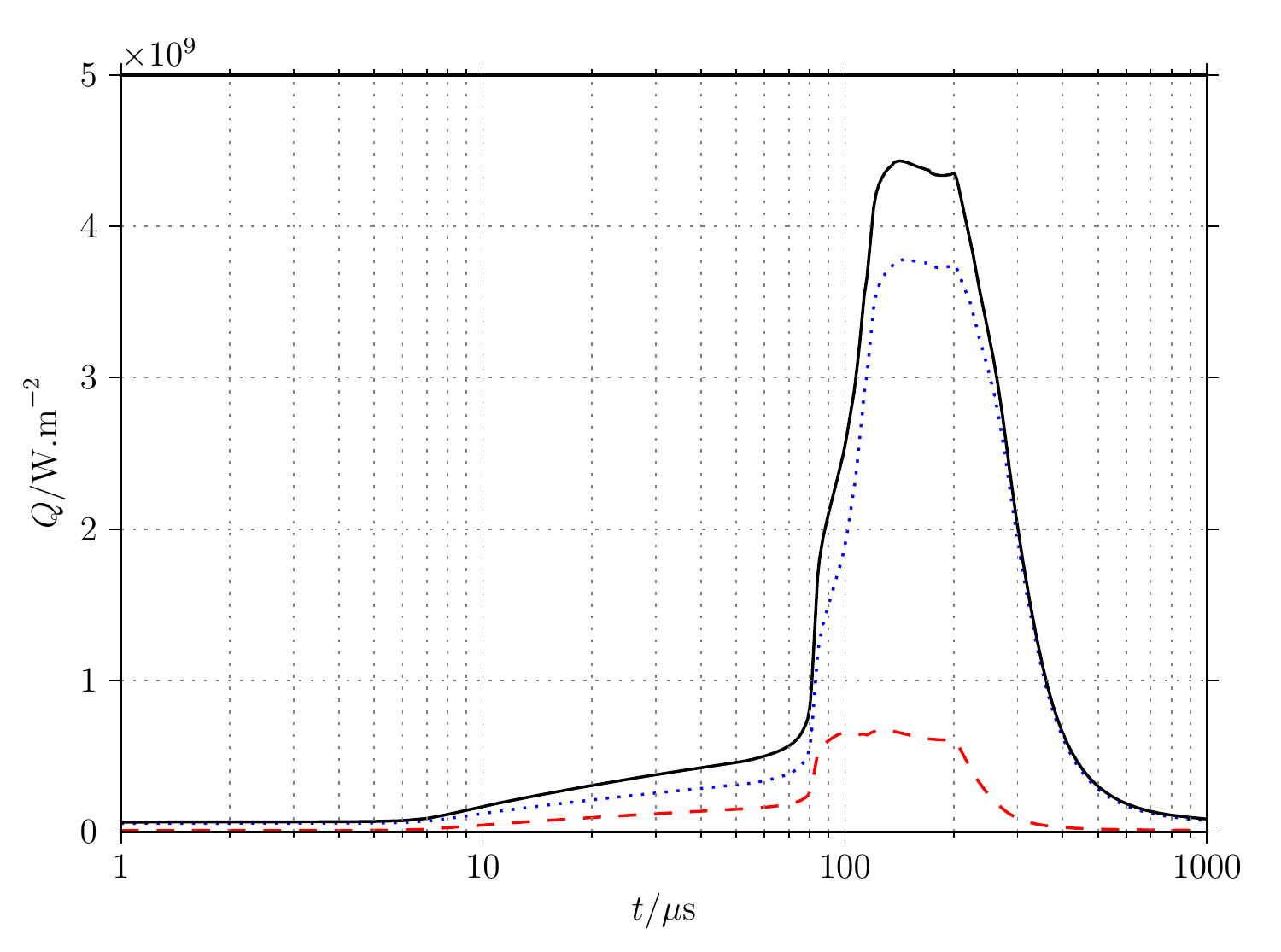}\hfill{}\includegraphics[width=0.49\textwidth]{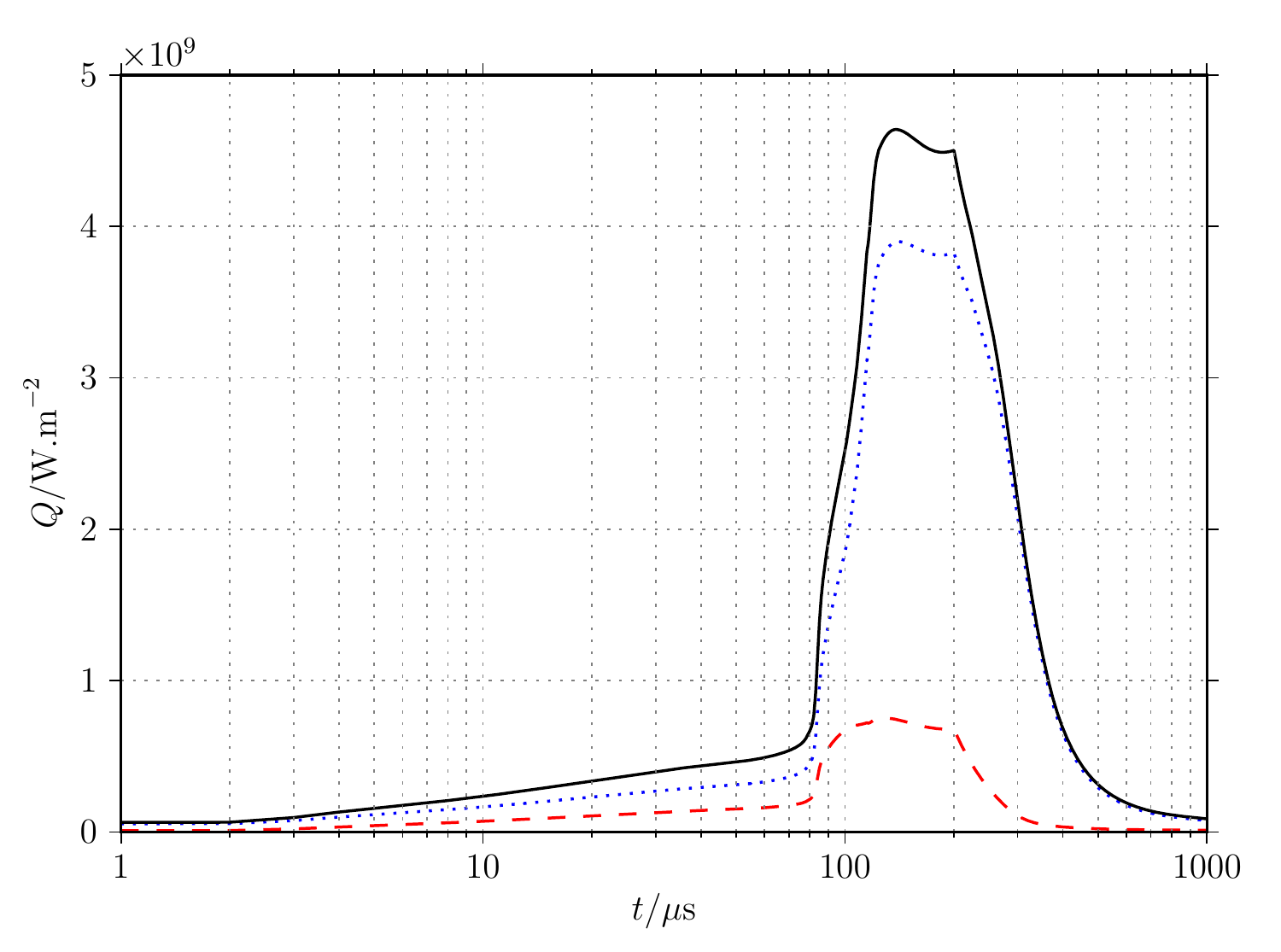}

\caption{Target heat-flux for simulations using non-local electron heat-flux
(left) and unlimited diffusive electron heat-flux (right): electron
component (red, dashed), ion component (blue, dotted) and total (black,
solid). N.B.~this is the \emph{target} heat-flux (at the material
surface) and not the \emph{sheath-edge} heat-flux that is used as
the boundary condition on the fluid model, i.e~the electron heat-flux
has been shifted down and the ion heat-flux shifted up by the sheath
potential. The heat-flux has not, however, been adjusted for the incidence
angle at the target, so if the angle were, for instance, $6^{\circ}$
the true heat-flux at the material surface would be smaller by a factor
of $\sim0.1$.\label{fig:target-heat-fluxes}}

\end{figure*}

For our simulations we model the SOL as a flux tube of fixed width,
with flat cross-field profiles, and use the one-dimensional fluid
equations, assuming quasineutrality and ambipolarity,\begin{align}
\frac{dn}{dt}+n\nabla_{\|}V_{\|} & =S_{n}\\
m_{i}n\frac{dV_{\|}}{dt} & =-\nabla_{\|}\left(nT_{i}\right)-\nabla_{\|}\pi_{i\|}-\nabla_{\|}\left(nT_{e}\right)\nonumber \\
 & \quad-m_{i}V_{\|}S_{n}\\
\frac{3}{2}n\frac{dT_{i}}{dt}+nT_{i}\nabla_{\|}V_{\|} & =-\nabla_{\|}q_{i\|}-\pi_{i\|}\nabla_{\|}V_{\|}\nonumber \\
 & \quad+\frac{3n\, m_{e}}{m_{i}\tau_{ei}}\left(T_{e}-T_{i}\right)+S_{E}-\frac{3}{2}T_{i}S_{n}\\
\frac{3}{2}n\frac{dT_{e}}{dt}+nT_{e}\nabla_{\|}V_{\|} & =-\nabla_{\|}q_{e\|}+\frac{3n\, m_{e}}{m_{i}\tau_{ei}}\left(T_{i}-T_{e}\right)\nonumber \\
 & \quad+S_{E}-\frac{3}{2}T_{e}S_{n}\label{eq:Te_equation}\end{align}
where $\frac{d}{dt}\equiv\left(\frac{\partial}{\partial t}+V_{\|}\nabla_{\|}\right)$,
$S_{n}$ is the particle source, $S_{E}$ is the energy source (which
is taken to be equal for electrons and ions) and $\tau_{ei}$ is the
electron-ion collision time. The equations are closed by specifying
$q_{e\|}$, $q_{i\|}$ and $\pi_{i\|}$:
\begin{itemize}
\item $q_{e\|}$ is given either by the non-local model described above,
by the Braginskii diffusion operator $q_{d,e}=-3.16\frac{n_{e}T_{e}\tau_{ei}}{m_{e}}\nabla_{\|}T_{e}$,
or by the flux-limited diffusion operator\begin{align}
q_{dl,e} & =\left(\frac{1}{q_{d,e}}+\frac{1}{\alpha_{e}q_{fs,e}}\right)^{-1}\label{eq:electron-flux-limiter}\end{align}
where the free-streaming heat-flux is $\left|q_{fs,e}\right|=nT_{e}v_{Te}$
and $\alpha_{e}$ is a parameter to be set.
\item $q_{i\|}$ is calculated using a flux-limited diffusion operator\begin{align}
q_{i\|} & =\left(\frac{1}{q_{d,i}}+\frac{1}{\alpha_{i}q_{fs,i}}\right)^{-1}\end{align}
where the Braginskii diffusion operator is $q_{d,i}=-3.9\frac{nT_{i}\sqrt{2}\tau_{ii}}{m_{i}}\nabla_{\|}T_{i}$,
the free-streaming heat-flux is $\left|q_{fs,i}\right|=nT_{i}v_{Ti}$
and $\alpha_{i}$ is a parameter to be set.
\item $\pi_{i\|}$ is given by a limited form of the Braginskii viscosity\begin{align}
\pi_{i\|} & =\left(\frac{1}{\pi_{0}}\pm\frac{1}{bnT_{i}}\right)^{-1}\equiv\frac{b\pi_{0}}{\left(\frac{\left|\pi_{0}\right|}{nT_{i}}+b\right)}\end{align}
where $\pi_{0}=-\frac{4}{3}\times0.96nT_{i}\sqrt{2}\tau_{ii}\nabla_{\|}V_{\|}$
and $b$ is a parameter to be set.
\end{itemize}

\subsection{Boundary Conditions\label{sub:Boundary-Conditions}}

In the SOL the boundary conditions are a critically important part
of the dynamics. Here we wish to impose the sheath-edge fluid boundary
conditions\citep{stangeby2000plasma}\begin{align}
V_{se} & =\pm c_{s}=\pm\sqrt{\frac{T_{e}+\gamma T_{i}}{m_{i}}}\label{eq:V_bc}\\
q_{se,e} & =\left(2T_{e}+\left|e\phi_{s}\right|\right)nc_{s}\label{eq:q_bc}\end{align}
where $e\phi_{s}=0.5T_{e}\ln\left(2\pi\frac{m_{e}}{m_{i}}\left(1+\frac{\gamma T_{i}}{T_{e}}\right)\right)$
is the sheath potential and $\gamma=3$, representing an approximately
collisionless sheath.

\eqref{eq:V_bc} and \eqref{eq:q_bc} under-determine the boundary
values of the non-Maxwellian moments $\hat{n}^{A}$. Since our aim
here is to develop computationally efficient methods to improve fluid
turbulence models, we simply choose the boundary moments so as to
impose \eqref{eq:q_bc} on the heat-flux and leave the remaining moments
unchanged, by adding a contribution proportional to $\mat{(W^{-1})}A{(1,1)}$.

There is an additional complication, that in a hot enough plasma (which
is found in the simulations described below) the collision length
may be long enough that the transients originating at one boundary
do not decay to zero by the time they reach the other. This means
that the boundary conditions cannot be set locally: one requires information
from both boundaries to set them consistently.

\section{Simulations\label{sec:Simulations}}

\begin{figure*}[t]
\includegraphics[width=0.49\textwidth]{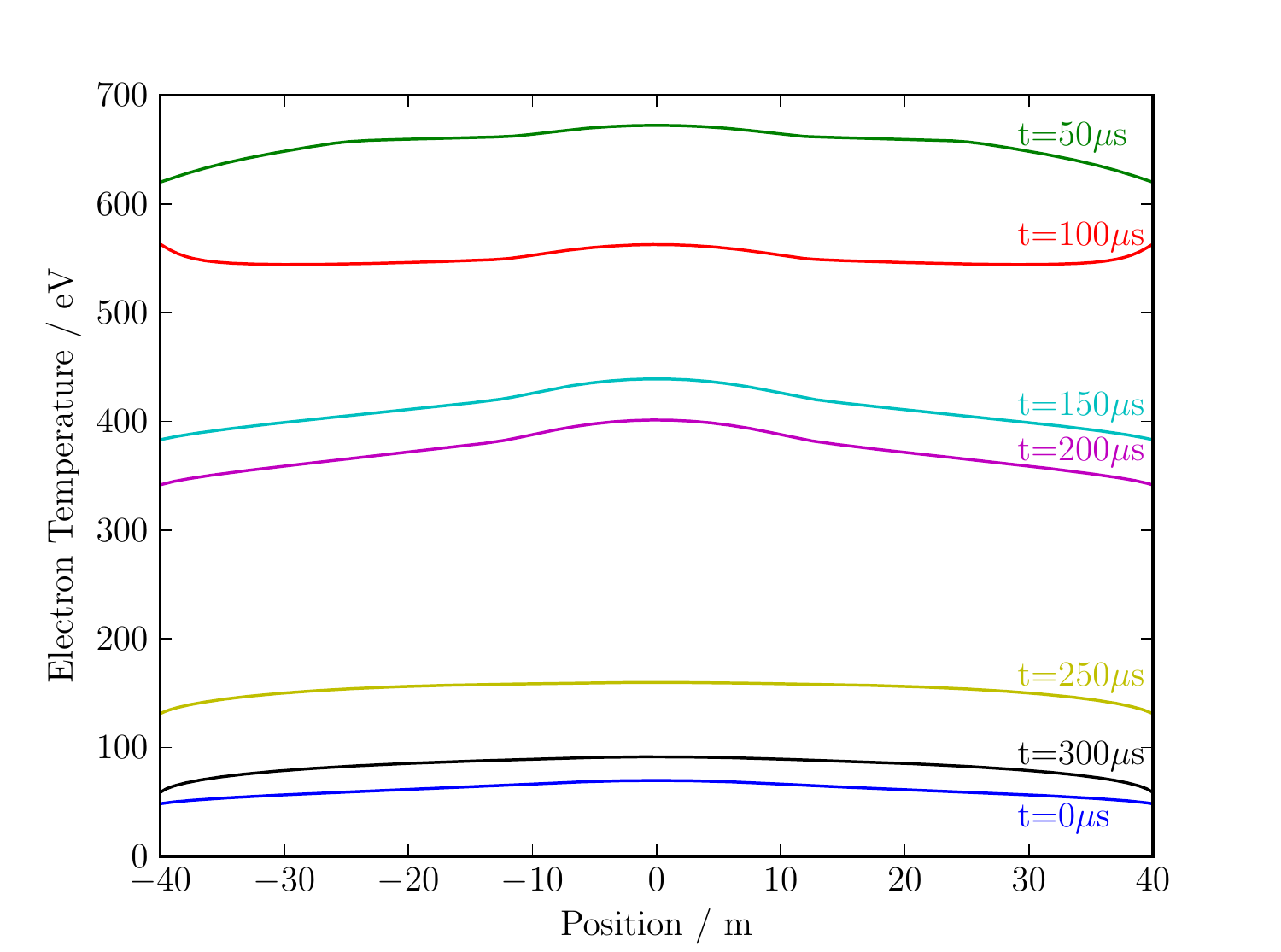}\hfill{}\includegraphics[width=0.49\textwidth]{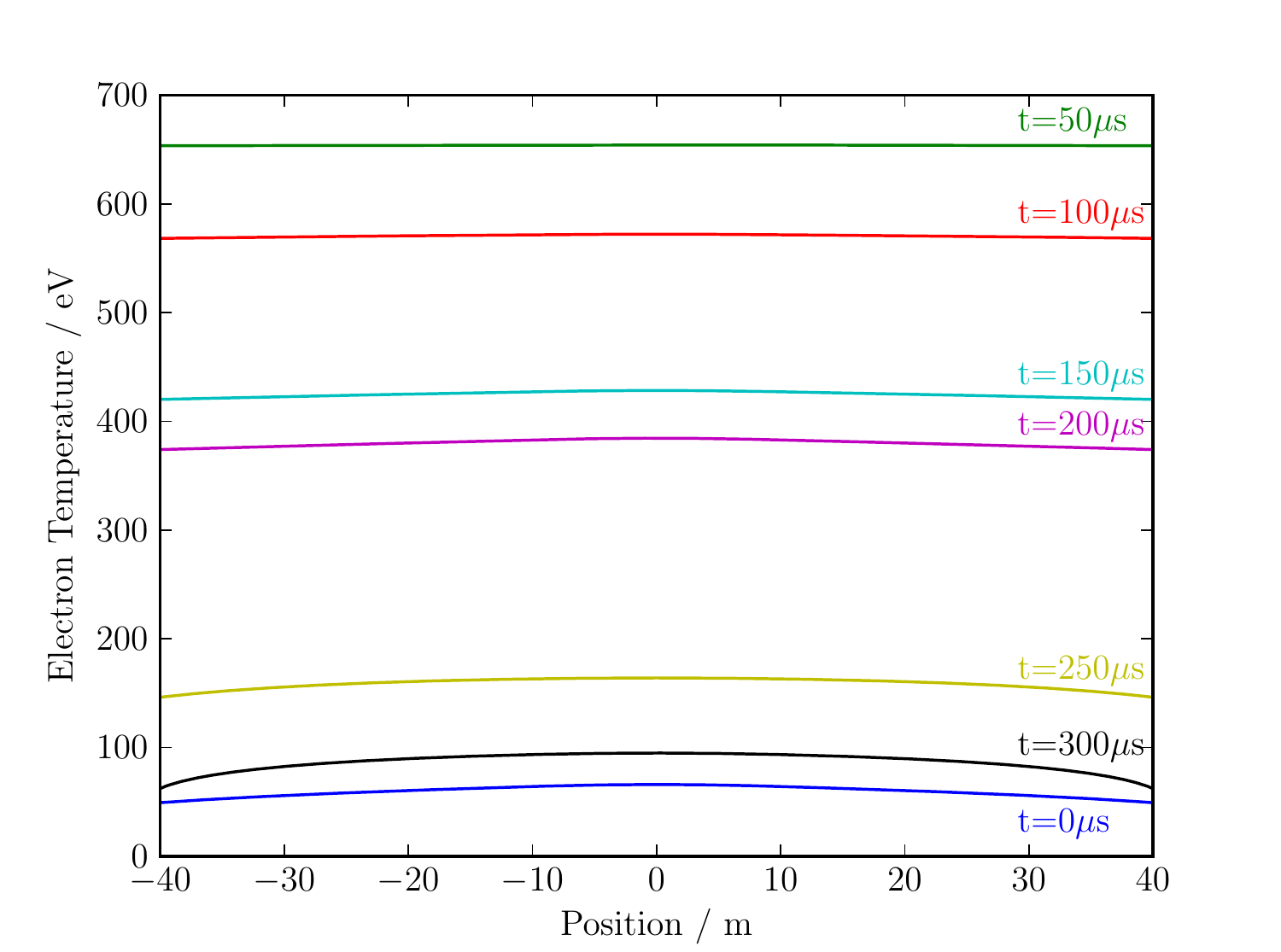}

\caption{Electron temperature profiles at various times during and after a
$200\mu\text{s}$ ELM using the non-local heat-flux model (left) and
a diffusive heat-flux model with a flux limiter of 0.4 (right).\label{fig:Te_profiles}}

\end{figure*}
We conducted simulations to compare the non-local electron heat-flux
model with the usual diffusive model. The test case chosen is that
used by \citep{havlivckova2012comparison}, which is representative
of the JET SOL. An ELM crash is simulated by introducing transient
sources of heat and particles with duration 200$\mu$s whose spatial
distributions are cosines about the centre of the simulation domain.
The total ELM energy is 0.4MJ, which is distributed evenly between
the ions and electrons, both at the pedestal temperature $T_{ped}=1.5\text{keV}$.
The source region of the SOL is taken to have a width of 10cm, a radius
of 3m, a poloidal extent of 2.6m and a field line pitch of $6^{\circ}$,
giving a volume of $4.9\text{m}^{3}$ and a parallel extent of $2.6\text{m}/\sin\left(6^{\circ}\right)\approx25\text{m}$.
Explicitly the source terms are \begin{align}
S_{ne}=S_{ni} & =\begin{cases}
\sigma_{n}\cos\left(\frac{\pi\ell}{L_{s}}\right) & \left|\ell\right|<\frac{L_{s}}{2}\\
0 & \text{else}\end{cases}\end{align}
\begin{align}
S_{Ee}=S_{Ei} & =\begin{cases}
\sigma_{E}\cos\left(\frac{\pi\ell}{L_{s}}\right) & \left|\ell\right|<\frac{L_{s}}{2}\\
0 & \text{else}\end{cases}\end{align}
where $\ell$ is again the co-ordinate parallel to the magnetic field
and $\ell=0$ at the centre of the domain, $L_{s}=25\text{m}$ is
the length of the source, $\sigma_{n}\approx9.1\times10^{23}\text{m}^{-3}\text{s}^{-1}$
and $\sigma_{E}\approx3.3\times10^{8}\text{J}\text{m}^{-3}\text{s}^{-1}$.
The total length of the domain is 80m. The initial profiles are found
by allowing the simulation to relax to a steady state with sources
that have similar form but smaller amplitude ($\sigma_{n}\approx7.4\times10^{22}\text{m}^{-3}\text{s}^{-1}$
and $\sigma_{E}\approx4.5\times10^{6}\text{J}\text{m}^{-3}\text{s}^{-1}$).
The ion heat-flux limiter used was $\alpha_{i}=0.1$ and the ion viscosity
limiter was $b=0.5$, as indicated by inter-ELM PIC simulations\citep{Tskhakaya2008}.
There are 256 uniformly distributed spatial grid points and a staggered
grid is used, with fluxes evaluated at the cell edges.

\begin{figure*}[t]
\includegraphics[width=0.49\textwidth]{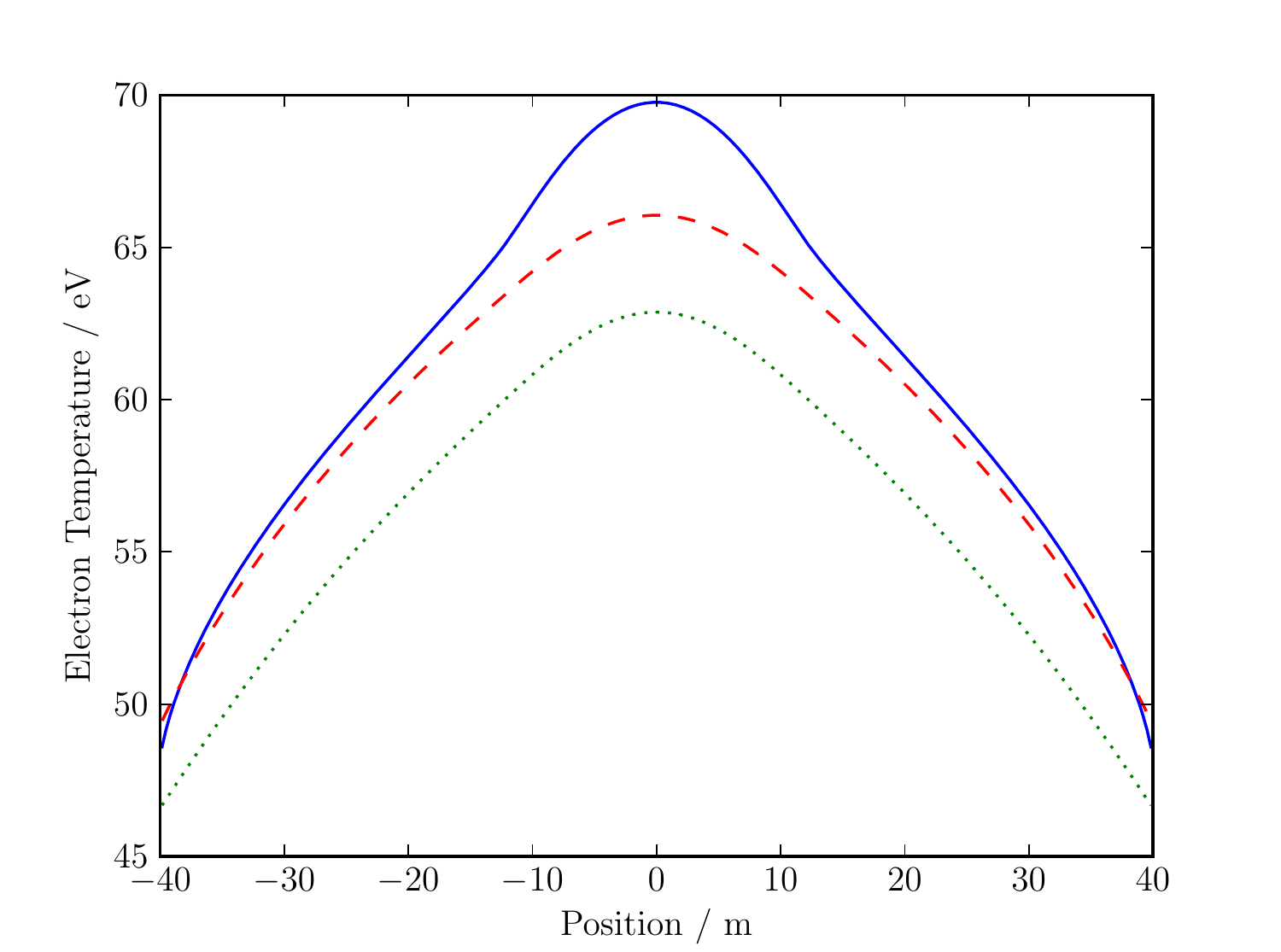}\hfill{}\includegraphics[width=0.49\textwidth]{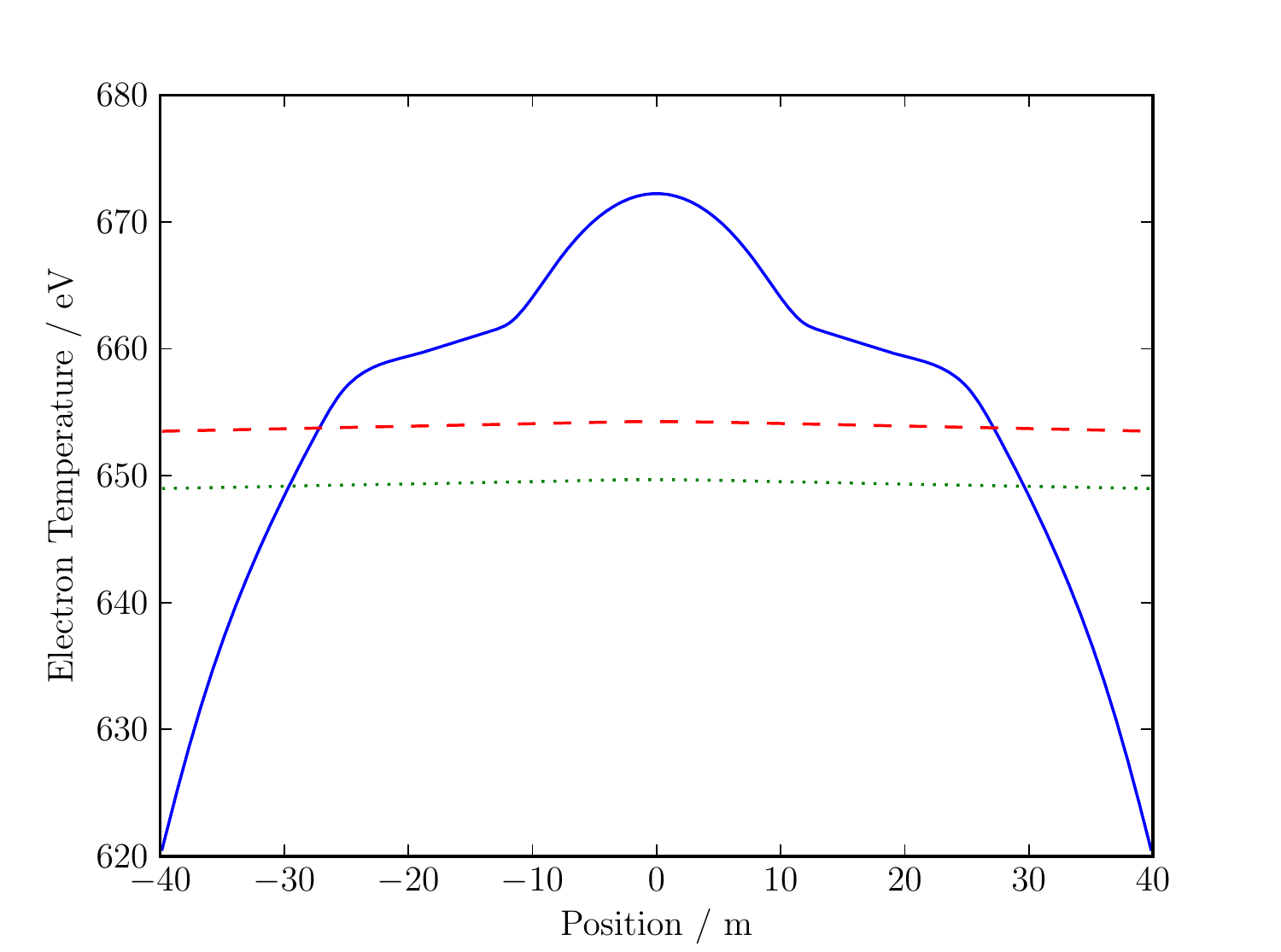}

\caption{Comparison of the electron temperature profiles from the non-local
(solid, blue) and diffusive models with a flux limiter of 0.4 (dashed,
red) and without flux limiter (dotted, green): on the left at $t=0\mu\text{s}$
(before the ELM) and on the right at $t=50\mu\text{s}$ (during the
ELM).\label{fig:Te_snapshot_comparisons}}

\end{figure*}

\begin{figure*}
\begin{minipage}[t]{0.49\textwidth}%
\includegraphics[width=1\textwidth]{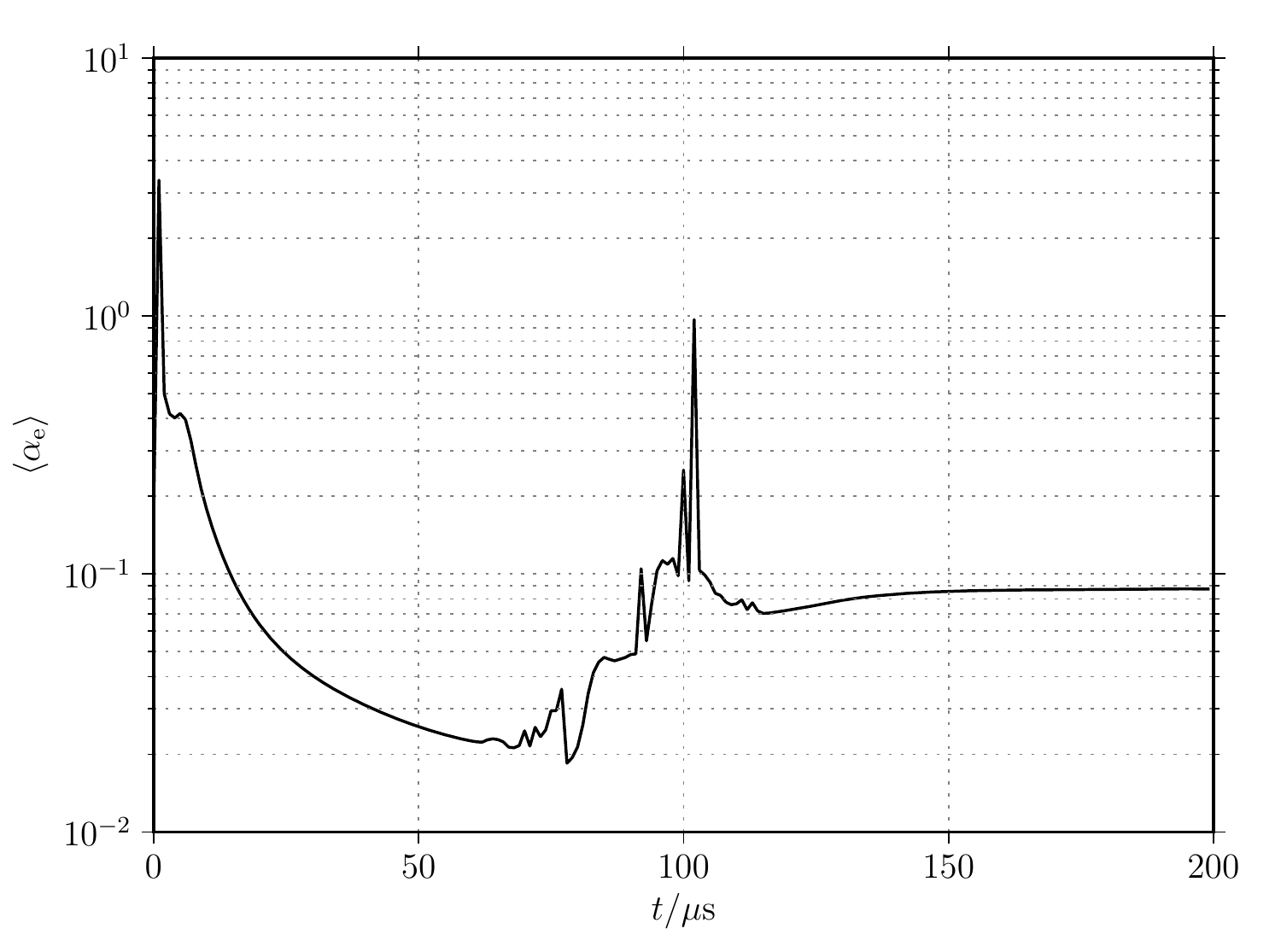}\caption{Spatially averaged value of the flux limiter, $\alpha_{e}$, that
would match the value of the non-local heat-flux.\label{fig:Effective_alphae}}
\end{minipage}\hfill{}%
\begin{minipage}[t]{0.49\textwidth}%
\includegraphics[width=1\textwidth]{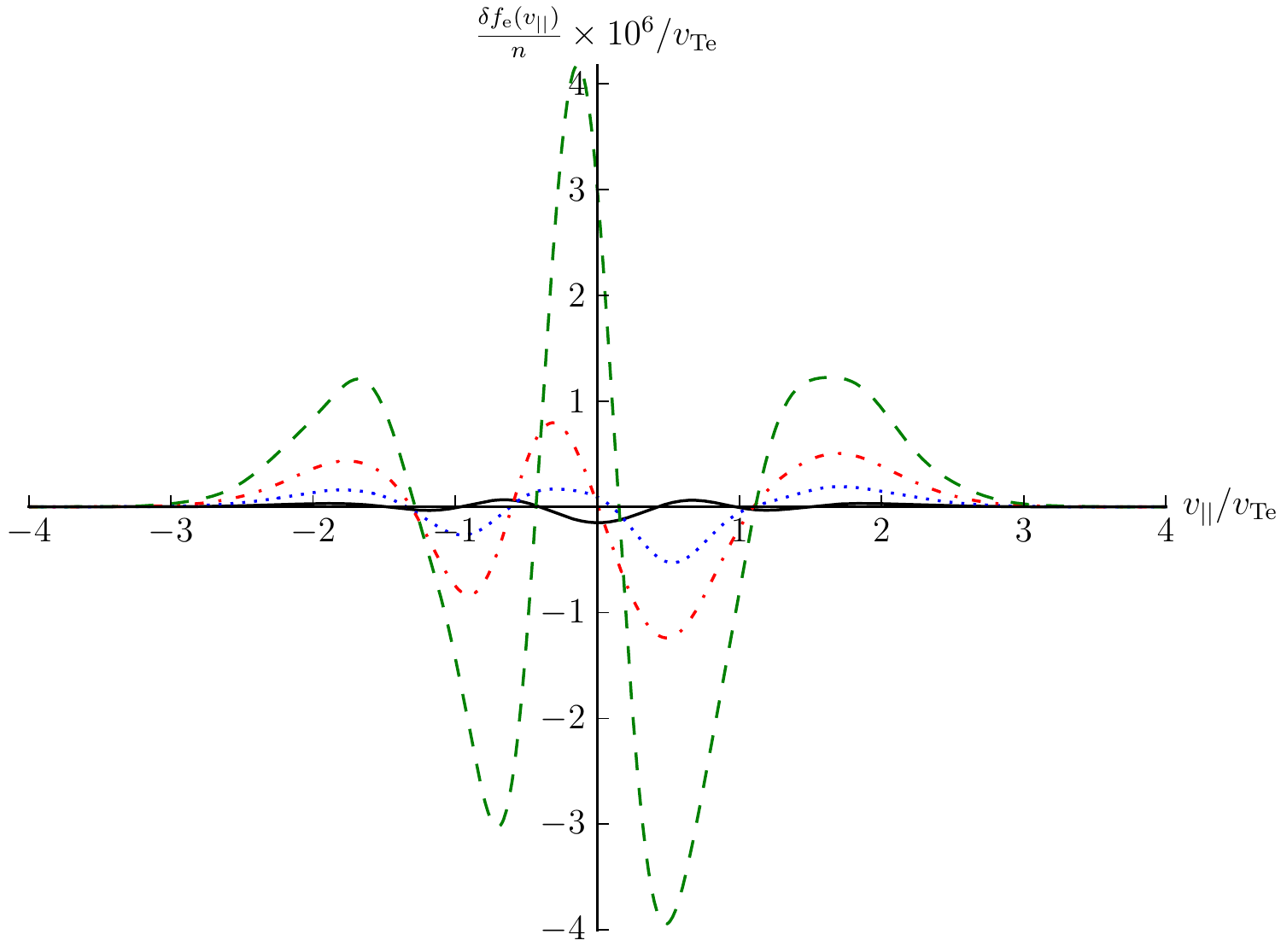}\caption{Non-Maxwellian part of the electron distribution function, integrated
over the perpendicular velocity and normalized by the density, with
velocities normalized by the thermal speed, at $t=50\mu\textrm{s}$
and positions $\ell=0.2\textrm{m}$ (near the mid-point; black, solid),
$\ell=13.3\textrm{m}$ (blue, dotted), $\ell=26.8\textrm{m}$ (red,
dash-dotted) and $\ell=40\textrm{m}$ (at the boundary; green, dashed).\label{fig:delta-f}}
\end{minipage}
\end{figure*}

It has been noted before\citep{pittsJET_ELM,havlivckova2012comparison}
that the target heat-flux is not strongly affected by kinetic effects,
the most important factor being the ions whose transport is mostly
convective and therefore well described by fluid models. It is not
surprising then that the target heat-flux is similar for the non-local
and the diffusive heat-flux models, as shown in Figure \ref{fig:target-heat-fluxes}.
One might have thought that using a model which includes kinetic effects
on the electron heat-flux would show the rapid response due to fast
electrons that is seen in PIC simulations\citep{havlivckova2012comparison}.
This is not the case here because both source terms and boundary conditions
are given in the fluid picture and do not include kinetic corrections.
In order to give a description more fully in agreement with the PIC
results some modification of the fluid boundary conditions would also
be necessary: in particular the sheath electron heat transmission
coefficient shows very strong variation through an ELM\citep{Tskhakaya2008}.
One proposal to achieve this is to use sheath coefficients corresponding
to the response to a collisionless Maxwellian wavepacket\citep{fundamenski2006},
but this relies on knowledge of the source to derive an explicit time
dependence. For three-dimensional simulations where the source on
a particular field line is not known in advance, a method which depends
only on the fields being evolved is needed. The non-local calculation
here depends only on the fluid variables, but derives from them extra
information about the non-Maxwellian part of the electron distribution
function. It would be interesting to investigate whether this can
be put to use in the construction of kinetically-corrected boundary
conditions. The conclusion of this is that for one-dimensional studies
the non-local electron heat-flux does not by itself offer much improvement
with respect to modelling of machine-relevant parameters, which are
dominated by the ion dynamics.

The value of the non-local model lies instead in aspects which will
be relevant when coupled in to three-dimensional simulations. As shown
in Figures \ref{fig:Te_profiles} and \ref{fig:Te_snapshot_comparisons}
the non-local model has a strong effect on the shape of the electron
temperature profiles. The qualitative difference between this non-local
model and a flux limiter approach is clear. Before the ELM, a flux
limiter with $\alpha_{e}=0.4$ is in fairly good agreement with the
non-local model, but during the ELM, when the electron temperature
is much higher and the collision length is much longer, the same flux
limiter cannot sustain temperature gradients of a similar order of
magnitude as the non-local model. To do so a much smaller flux limiter
would be needed, for example at $t=50\mu\text{s}$ a limiter with
$\alpha_{e}\sim0.025$ would be required (see Figure \ref{fig:Effective_alphae}).
A flux limiter cannot be set which is appropriate for the whole range
of conditions. This is in agreement with PIC simulations of the SOL\citep{Tskhakaya2008},
which show that to capture the kinetic effects the parameter of a
flux limiter model would need to vary both in space and time. In contrast
the non-local heat-flux can cover a much broader range of conditions.
A PIC simulation of the same sort of ELM (0.4MJ, $200\mu\text{s}$)
as we have simulated here shows\citep[Fig.\ 7]{Tskhakaya2008} that
the value of the electron heat-flux limiter that would match the kinetic
simulation becomes much smaller than the pre-ELM value after $50\mu\text{s}$,
showing that in the kinetic simulation a much steeper temperature
gradient is needed to drive the heat-flux from the source than would
be the case with a flux limiter set to the pre-ELM value. Figure \ref{fig:Effective_alphae}
shows a similar plot from our non-local heat-flux simulation: the
spatial average, $\langle\alpha_{e}\rangle$, of the flux limiter
parameter (here un-normalized, unlike \citep[Fig.\ 7]{Tskhakaya2008})
that would be needed at each point for the diffusion operator \eqref{eq:electron-flux-limiter}
to match the non-local heat-flux. In both cases the minimum value
reached by $\langle\alpha_{e}\rangle$ is similar. \citep[Fig.\ 6]{Tskhakaya2008}
shows that the kinetic correction to the electron sheath heat transmission
coefficient, $\gamma_{e}$, passes through 1 (i.e.~no correction)
at about $60\mu\text{s}$, which happens to be the time when $\langle\alpha_{e}\rangle$
reaches its minimum value. This is why our non-local fluid simulation,
which does not include any kinetic corrections to the fluid boundary
conditions, is able to find the same value. After $60\mu\text{s}$,
the fluid heat transmission coefficient is larger than the kinetic
one, and so the electrons are cooler in the fluid model, with a corresponding
decrease in collision length and increase in $\langle\alpha_{e}\rangle$.
This reemphasizes the importance of developing a better model of the
sheath boundary conditions for fluid simulations of the SOL.

\begin{figure*}[t]
\includegraphics[width=0.49\textwidth]{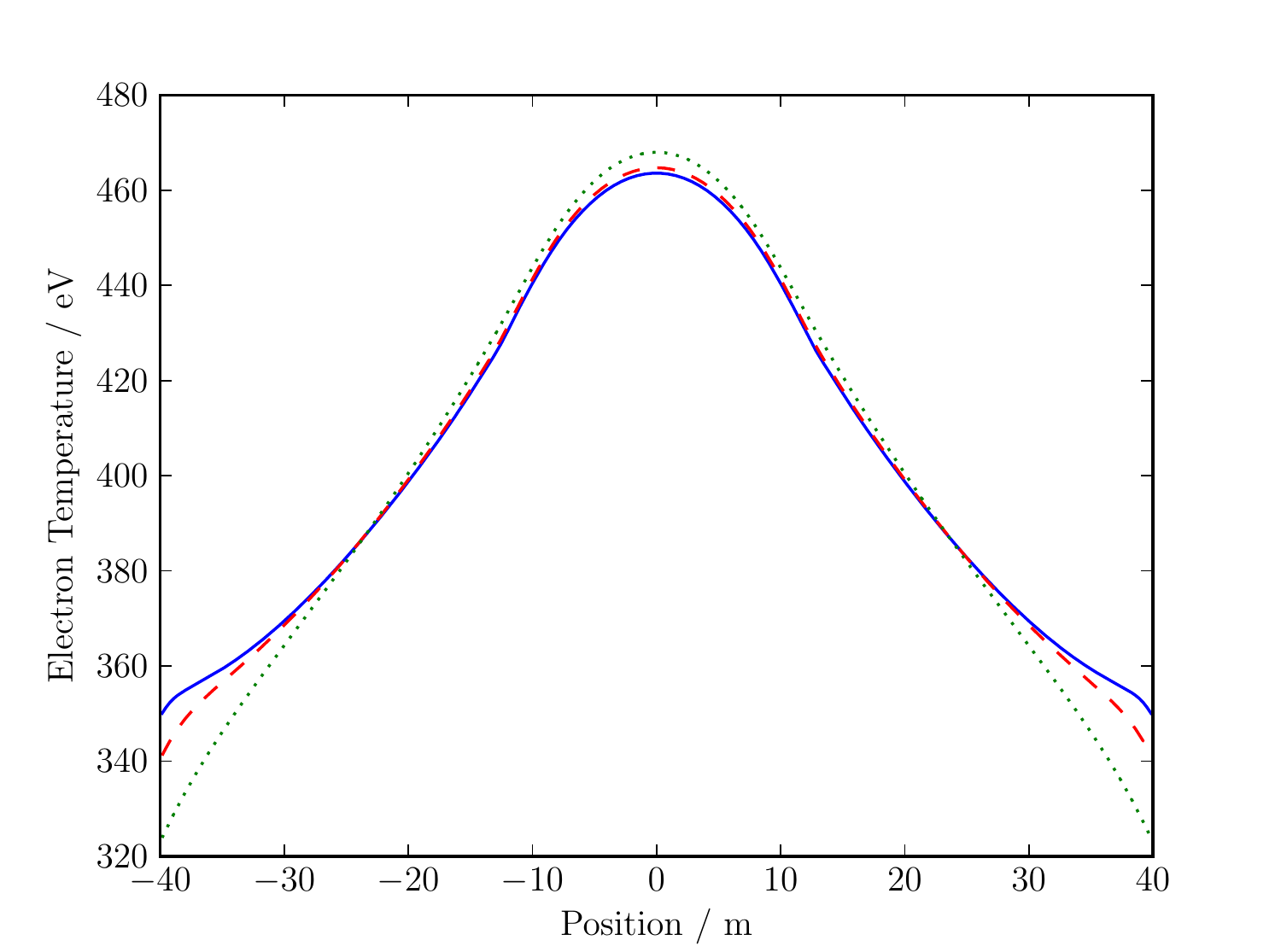}\hfill{}\includegraphics[width=0.49\textwidth]{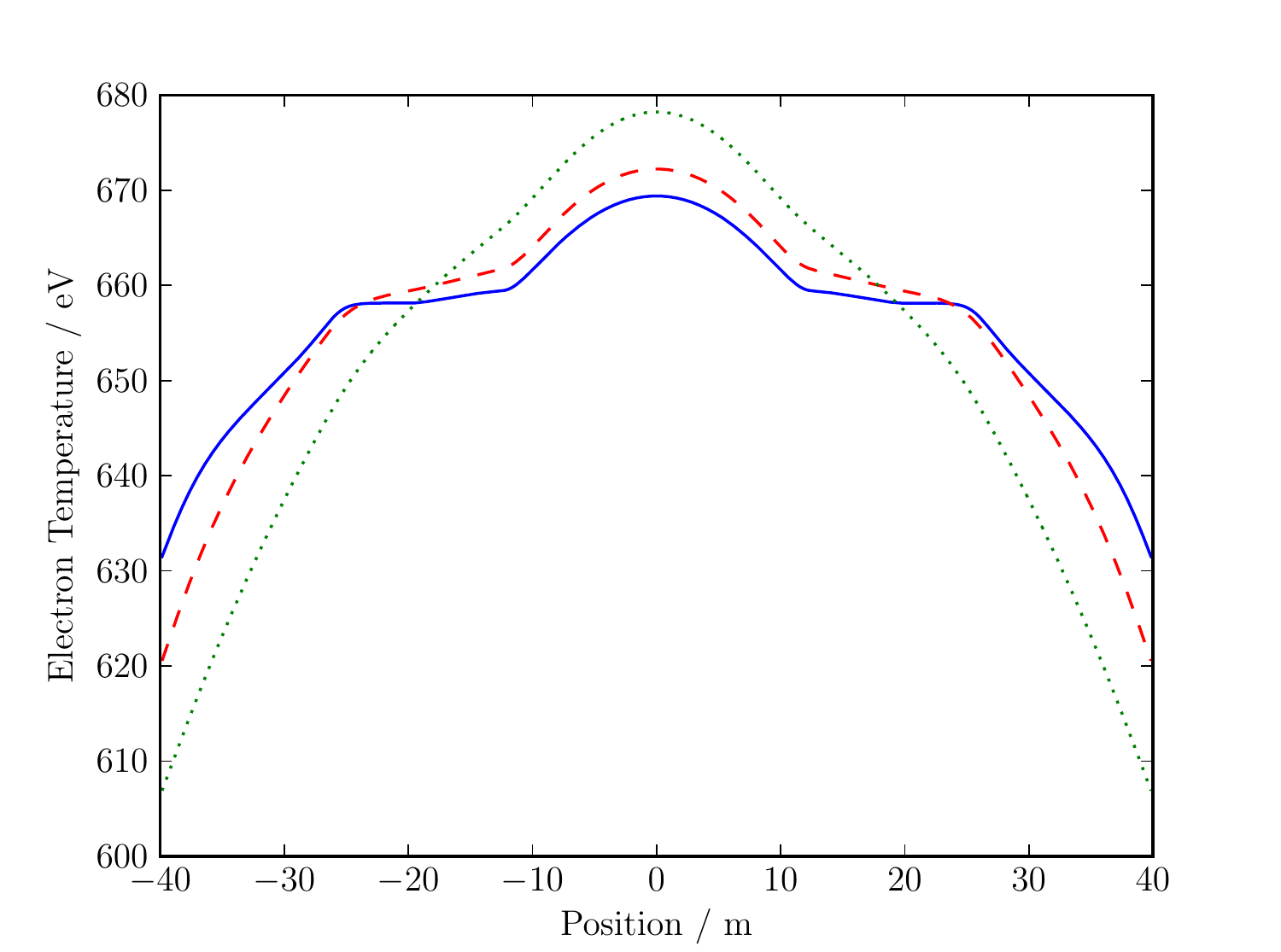}

\caption{Comparison of the electron temperature profiles for 100 (dotted, green),
400 (dashed, red) and 1600 (solid, blue) moment heat-flux models:
on the left at $t=15\mu\text{s}$ (as the 100 moment model begins
to diverge) and on the right at $t=50\mu\text{s}$ (near the peak
electron temperature).\label{fig:Te_snapshot_comparisons_moments}}

\end{figure*}

It is possible to reconstruct the electron distribution function from
this non-local approach. As an example Figure \ref{fig:delta-f} shows
the non-Maxwellian part of the distribution function, integrated over
perpendicular velocity and normalized by the density, $n^{-1}\negthickspace\int d^{2}v_{\perp}\delta f_{e}$,
for several points at $t=50\mu\textrm{s}$. The extent to which this
distribution function can capture the structure found in a fully kinetic
simulation remains to be investigated.

One might notice that the $t=100\mu\text{s}$ temperature profile
from the non-local model in Figure \ref{fig:Te_profiles} looks odd:
it increases near the boundaries instead of decreasing monotonically
from the centre of the domain. The reason is that as the density from
the ELM reaches the boundary it causes a temporary glitch while the
boundary conditions adjust, with a very steep velocity gradient appearing
near the boundary. This causes the convective term in \eqref{eq:Te_equation}
to heat the electrons, resulting in the upturn in temperature. The
diffusive model allows enough heat-flux to conduct away this extra
heat, but the non-local model does not, which is why this behaviour
is noticeable only in the non-local model. The reason for the glitch
may be the way the ion viscosity is calculated. A viscosity limiter
is needed since without it the viscous heating would increase the
temperature of the SOL plasma above the temperature of the source,
which is clearly unphysical. On the other hand without some viscosity
the velocity of the plasma would increase to above the sound speed.
This suggests perhaps that a better method of calculating the viscosity
may be useful. It is possible to calculate the ion viscosity in an
analogous manner to the calculation of the electron heat-flux used
here\citep{ji2009moment}. The neglect of time derivatives used in
this approach is obviously not strictly valid for ions when ELMs occur:
the plasma does then evolve significantly on ion timescales. Nevertheless
the calculation would be valid both before the ELM begins and in the
steady state reached if the ELM continues long enough. It might therefore
be reasonably hoped that, while not quantitatively accurate, it would
interpolate between the two states in a qualitatively better way than
a limiter which can be set only for one or the other state.

\subsection{Model Convergence\label{sub:Model-Convergence}}

In the non-local model one can increase the number of moments used;
the effect is to extend the validity of the model to longer collision
lengths\citep{ji:022312}, though at the cost of computational speed.
The simulations reported here used $L=20$, $K=20$ (400 moments).
To verify that this number is appropriate, we ran simulations with
fewer ($L=10$, $K=10$; 100 moments) and more ($L=40$, $K=40$;
1600 moments) moments. As shown in Figure \ref{fig:Te_snapshot_comparisons_moments},
the 400 and 1600 moment simulations agree well across the whole range
of temperatures in this simulation; they do differ somewhat in the
size of their response to the {}`glitch' mentioned above but since
that is unphysical this difference is not relevant. At the highest
temperatures (and hence longest collision lengths) the 100 moment
version gives significantly different results.

\section{Implementation\label{sec:Numerical-Performance}}

\begin{figure*}[t]
\begin{minipage}[t]{0.48\textwidth}%
\includegraphics[width=1\textwidth]{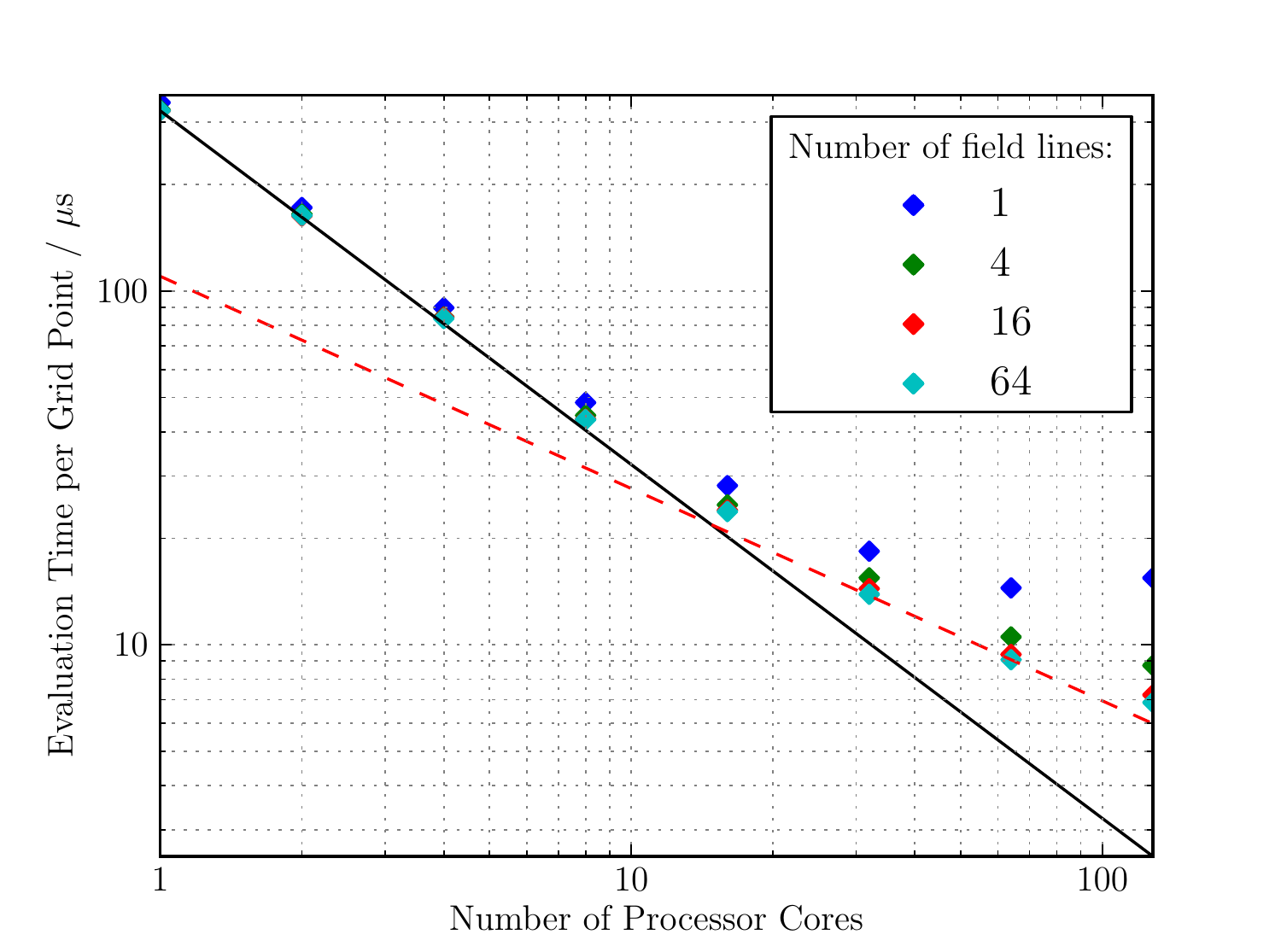}

\caption{Scaling performance of the non-local heat-flux computation for 400
moments on a 256 point grid on HPC-FF. Extra field lines allow more
efficient overlapping of computations, improving the scaling compared
to the single field line case. The black, solid line shows linear
scaling and the red, dashed line shows $n^{-0.6}$ scaling.\label{fig:scaling_performance}}
\end{minipage}\hfill{}%
\begin{minipage}[t]{0.48\textwidth}%
\includegraphics[width=1\textwidth]{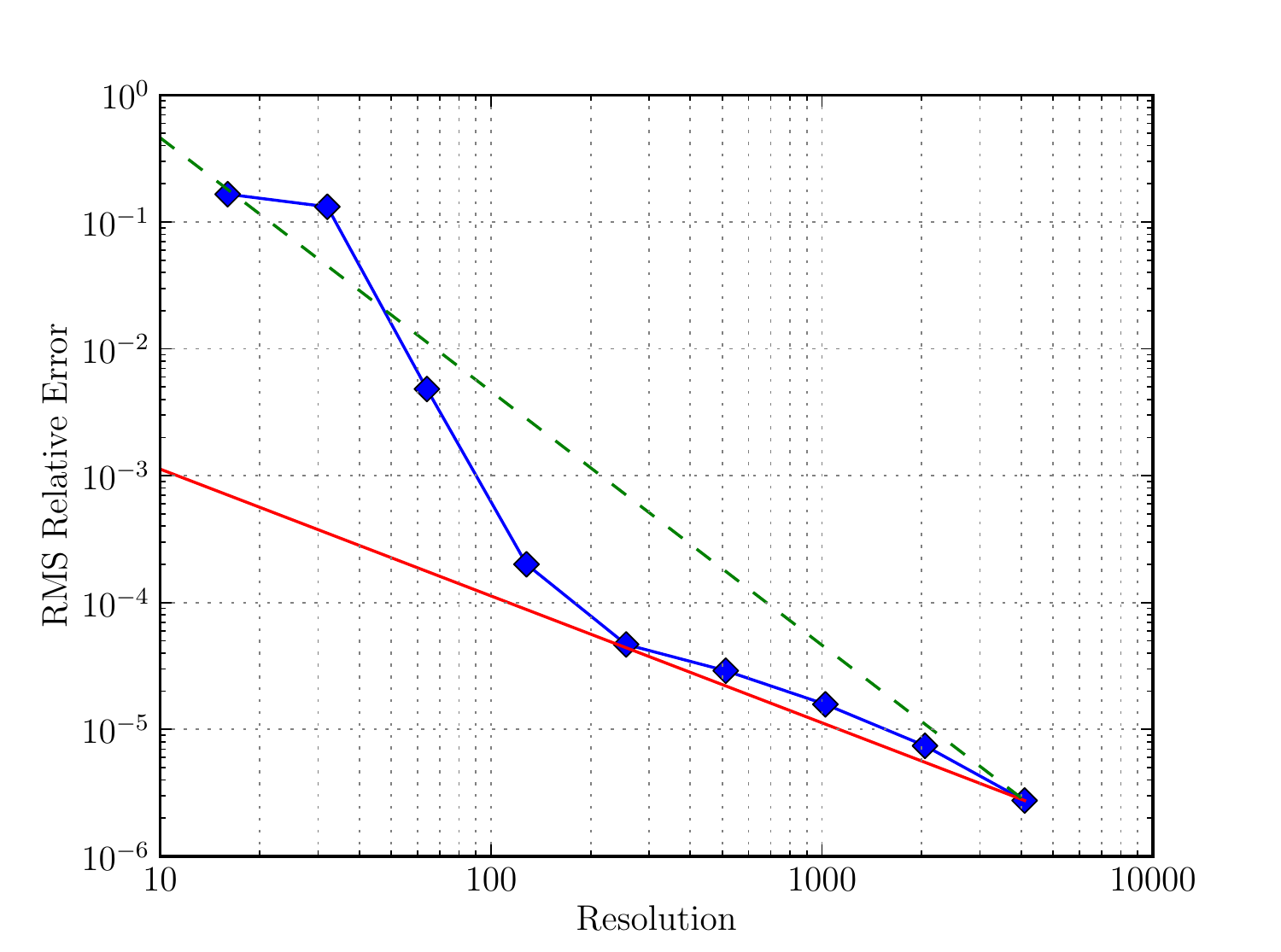}

\caption{Root-mean-square relative error compared to calculation with 9192
grid points. The red, solid line is $\propto n^{-1}$ and the green,
dashed line is $\propto n^{-2}$.\label{fig:spatial_convergence}}
\end{minipage}
\end{figure*}

The computation of the non-local heat-flux requires many integrals
to be calculated: one per moment per field line. Even though the integrand
depends on the point of evaluation, it is not necessary to compute
separate integrals for each point. The dependence, equation \eqref{eq:moment-solution},
is simple enough that integral at each point can be calculated from
that at the point either to the left or to the right (depending on
the sign of the eigenvalue associated with that integral) so that
the integrals at every point can be computed in a single loop over
the field line. The integrals for different moments and for different
field lines are independent, allowing efficient parallelization by
overlapping their computations on different processors to minimize
idle time. Since this is a one-dimensional, field-parallel calculation,
the scaling as the grid is divided in the other directions (i.e. radially
and toroidally) is perfect (i.e.~linear) since processors dealing
with separate field lines do not need to communicate. This favours
the division of a three-dimensional simulation grid as finely as possible
radially and toroidally before starting to distribute the grid points
on a single field line between different processors. However, other
operations have diametrically opposed requirements: BOUT++ does not
split the simulation grid at all in the toroidal direction in order
to allow efficient computation of toroidal Fourier transforms; and
perpendicular Laplacian inversions favour splitting of the grid parallel
to the field rather than radially\citep{Dudson2009}. Therefore it
is important that this heat-flux calculation scale efficiently when
the grid is split in the parallel direction, but with several field
lines on each processor. The sub-grid assigned to each processor will
have at least as many points as the toroidal size of the grid, and
likely several times that: this is the reason that Figure \ref{fig:scaling_performance}
shows the scaling for various numbers of field lines per processor,
even though the most efficient splitting for \emph{this} calculation
would be to have just one field line per processor. The maximum number
of field lines plotted is 64 since, for a grid with 256 parallel points,
the evaluation time per grid point for larger numbers of field lines
did not change significantly from that for 64. Figure \ref{fig:scaling_performance}
shows that scaling (for a 256 point grid) is near linear up to $16=\sqrt{256}$
processors, falling to $\sim n^{-0.6}$ above that as the communication
time becomes relatively more important.%
 The efficiency (for 64 field lines per processor) at 64 processors,
i.e.~4 parallel points per processor, is 56\% compared to using a
single processor.

The grid for a typical three-dimensional simulation in BOUT++ might
have 256 radial, 256 field-parallel and 256 toroidal points; dividing
the grid to put 4 radial and 4 parallel points on each processor would
mean using 4096 processors. From the point of view just of this parallel
heat-flux calculation this means 64 independent sets of 64 processors,
with each set having 1024 field lines. As the number of field lines
is greater than 64, the efficiency will still be 56\% and the total
evaluation time (for the heat-flux calculation) for the $256\times256\times256$
grid will then be $\sim2.4\text{s}$. Including this non-local heat-flux
in a three-dimensional simulation will not be detrimental to the scaling
performance of BOUT++, and so we anticipate good efficiency up to
several thousand processors, as has been previously demonstrated for
BOUT++ simulations\citep{Dudson2009}.

The run-time of the 400 moment simulation until the end of the $200\mu\textrm{s}$
duration of the ELM is about 7 minutes on 32 processors of HPC-FF
(with just one field line), so to do the same calculation on a three-dimensional
$256\times256\times256$ grid would take $\sim2.5$ days on 4096 processors.
This indicates that three-dimensional fluid simulations with non-local
parallel transport will be practicable. For comparison, \citep{Tskhakaya2012}
concluded that three-dimensional PIC simulations are beyond the reach
of current or near-future supercomputers, as they would require $\geq3\times10^{6}$
teraflops to run in one week. Indeed, even in one dimension, a single
run of the BIT1 PIC code takes upwards of 12 hours on 512 processors\citep{Tskhakaya2012},
which would make it challenging even to use the PIC code just for
parallel kinetic corrections in a three-dimensional fluid model, with
many thousands of field lines.

The spatial convergence of the calculation compared to a 9192 point
grid is shown in Figure \ref{fig:spatial_convergence}. 256 point
resolution has a root mean square relative error of $4.7\times10^{-5}$.
The input data for the test are some functions chosen to have gradients
comparable to the biggest ones found during the ELM simulations (modelled
on the $t=2\mu\text{s}$ slice).

\section{Conclusions\label{sec:Conclusions}}

By using a non-local operator to calculate the electron heat-flux,
we can account for some kinetic effects in fluid models, without needing
to set parameters by comparison with kinetic simulations or experiment.
This is especially beneficial when transients such as ELMs occur because
the non-local heat-flux can respond self-consistently to the changing
conditions in the plasma in a way that flux limiters cannot.

In one dimensional simulations, the effect on machine-relevant parameters,
such as the heat-flux at the divertor targets, is not significant
as their behaviour is dominated by the ion dynamics (Figure \ref{fig:target-heat-fluxes})
which, being largely convective, are well described by fluid models.

However, there is a substantial change in the shape of the electron
temperature profiles (Figures \ref{fig:Te_profiles} and \ref{fig:Te_snapshot_comparisons})
which will alter the drive of turbulence in three-dimensional simulations
and thus will affect, for example, the width of the strike-point on
the divertor. The principal advantage of the technique described here
is that it will allow self-consistent three-dimensional simulations:
since the sources of heat and particles on each field line would then
be determined by the cross-field transport, achieving a similar effect
through flux limiters set by comparison to one-dimensional PIC simulations
would be very challenging. 

The main limitation preventing a closer match of the electron dynamics
between these non-local fluid simulations and PIC simulations such
as those in \citep{Tskhakaya2008} seems to be the response of the
sheath-edge boundary conditions to kinetic effects during transients,
which have not yet been included in our fluid simulations.

The non-local heat-flux has been implemented in BOUT++ and parallelized
efficiently. It is ready to be included in three-dimensional simulations
of the SOL.

\subsection{Future work}

The next step in this work is to move on to three-dimensional turbulence
simulations of the SOL, to investigate the effects on turbulence and
perpendicular transport of the electron temperature gradients seen
here that are missed by local heat-fluxes currently used in SOL fluid
models.

In order to correctly describe realistic tokamak geometries, the model
should be extended to include the effects of $\nabla B$ on the heat-flux
calculation. The effect will be to add an additional drive term to
equation \eqref{eq:kinetic_equation} which, as it depends only on
the Maxwellian part of the distribution function, should add only
a modest amount of additional complication.

A detailed comparison to one-dimensional kinetic simulations should
be carried out to assess how much of the kinetic information can be
captured by the non-local model, for instance: how accurately and
up to what range in velocity space the deviation from a Maxwellian
(e.g.~that shown in \citep{Tskhakaya2012}) can be described.

In the non-local approach one has more information about the electron
distribution function than just the density and temperature. It may
be possible to use this information to improve the modelling of the
sheath and so find more accurate boundary conditions for SOL fluid
models. Whether by this method or another, finding a way to include
some kinetic corrections to the sheath transmission coefficients would
greatly improve the accuracy of fluid descriptions of parallel transport
in the SOL.

Since convection dominates over conduction for the ions, kinetic corrections
are less important; flux and viscosity limiters describe the behaviour
of the ions better than is the case for the electrons. There may however
be some scope for improvement in this area from application of a non-local
model to the ion dynamics also. The variation of the ion viscosity
limiter (inferred from a PIC simulation) through an ELM crash is not
as dramatic as that of the electron heat-flux limiter\citep{Tskhakaya2008}.
However, the value of the viscosity limiter does have a significant
effect on the dynamics, affecting in particular the speed reached
by the plasma and hence the time for the ELM to reach the target.
Though a non-local calculation on the same lines as that used here
for the electron heat-flux\citep{ji2009moment} is not guaranteed
to be valid for the ion dynamics due to the neglect of time derivatives,
it seems worth investigating whether it might be able to give a better
qualitative agreement with kinetic simulations and therefore be of
use for improving fluid turbulence simulations.

\section*{Acknowledgements}

We gratefully acknowledge EFDA support for the use of the HPC-FF system
through project FSNBOUT.

\setlength{\bibsep}{1pt}\bibliographystyle{unsrtnat}
\bibliography{references}

\end{document}